\documentclass[10pt, aps, prb, twocolumn, showpacs, amsmath, amssymb, amscd,
superscriptaddress, floatfix, longbibliography
]{revtex4-1}

\usepackage{braket}

\usepackage[version=4]{mhchem}
\usepackage[dvipsnames]{xcolor}
\usepackage{bbold}

\definecolor{first}{RGB}{0,153,230}
\definecolor{second}{RGB}{40,37,110}
\definecolor{third}{RGB}{242,56,20}

\usepackage[hidelinks]{hyperref}
\usepackage[capitalize]{cleveref}
\usepackage{csquotes}

\usepackage{diagbox}

\usepackage{bbold}

\usepackage{pifont}
\usepackage{tikz}
\usetikzlibrary{shapes,arrows, patterns, shadows,positioning, decorations,
    calc}
\usepackage{xcolor}
\usepackage{amsmath, amssymb}

\makeatletter
\def\convertto#1#2{\strip@pt\dimexpr #2*65536/\number\dimexpr 1#1}
\makeatother
\newcommand{\measurewidths}{{%
  \color{red}
  :::
  TEXT \convertto{in}{\the\textwidth} in
  :::
  COLUMN \convertto{in}{\the\columnwidth} in
  :::
  HEIGHT(ex) \convertto{pt}{1ex} pt
  :::
  WIDTH(em) \convertto{pt}{1em} pt
  :::
}}

\usepackage{makerobust}
\newcommand*\circled[1]{\tikz[baseline=(char.base)]{
    \node[shape=circle,draw,inner sep=1pt] (char) {#1};}}
\MakeRobustCommand\circled

\AtBeginDocument{%
    \newwrite\bibnotes
    \def\bibnotesext{Notes.bib}
    \immediate\openout\bibnotes=\jobname\bibnotesext
    \immediate\write\bibnotes{@CONTROL{REVTEX41Control}}
    \immediate\write\bibnotes{@CONTROL{%
    apsrev41Control,author="08",editor="1",pages="1",title="0",year="1"}}
     \if@filesw
     \immediate\write\@auxout{\string\citation{apsrev41Control}}%
    \fi
}%

\newcommand{\bvec}[1]{\boldsymbol #1}

\newcommand{\cre}[1]{\ensuremath{c^{\vphantom{\dagger}}_{#1}}}
\newcommand{\ann}[1]{\ensuremath{c^{\dagger}_{#1}}}

\newcommand{\sym}[2]{\ensuremath{\text{#1}_{#2}}}

\newcommand{\makeauthor}[2]{\newcommand{#1}[1]{{%
  \sffamily\color{#2}{%
    \bfseries\begingroup\escapechar=-1\edef\x{\endgroup\string#1}\x:%
  } ##1}}%
  \MakeRobustCommand#1}

\makeauthor{\jb}{ForestGreen}
\makeauthor{\sr}{blue}
\makeauthor{\mb}{red}

\newcommand{\up}{{\uparrow}}
\newcommand{\dw}{{\downarrow}}

\newcommand{\bs}[1]{\boldsymbol{#1}}

\newcommand{\eps}{{\varepsilon}}

\newcommand{\ie}{{\it i.e.},\ }
\def\eg{\emph{e.g.}\ }

\graphicspath{{./}{./plots/}}

\begin{document}

\title{Chern number landscape of spin-orbit coupled chiral superconductors}
\author{Matthew Bunney}
\thanks{These authors contributed equally.}
\affiliation{School of Physics, University of Melbourne, Parkville,
    VIC 3010, Australia}
\affiliation{Institute for Theoretical Solid State Physics,
    RWTH Aachen University, 52062 Aachen, Germany}

\author{Jacob Beyer}
\thanks{These authors contributed equally.}
\affiliation{School of Physics, University of Melbourne, Parkville,
    VIC 3010, Australia}
\affiliation{Institute for Theoretical Solid State Physics,
    RWTH Aachen University, 52062 Aachen, Germany}
\affiliation{JARA Fundamentals of Future Information Technology,
    52062 Aachen, Germany}
\affiliation{Institute for Theoretical Physics, University of Würzburg,
    Am Hubland, 97074 Würzburg, Germany}

\author{Ronny Thomale}
\affiliation{Institute for Theoretical Physics, University of Würzburg,
    Am Hubland, 97074 Würzburg, Germany}

\author{Carsten Honerkamp}
\affiliation{Institute for Theoretical Solid State Physics,
    RWTH Aachen University, 52062 Aachen, Germany}
\affiliation{JARA Fundamentals of Future Information Technology,
    52062 Aachen, Germany}

\author{Stephan Rachel}
\affiliation{School of Physics, University of Melbourne, Parkville,
    VIC 3010, Australia}


\begin{abstract}
Chiral superconductors are one of the predominant quantum electronic states of matter where topology, symmetry, and Fermiology intertwine. This is pushed to a new limit by further invoking the coupling between spin and charge degrees of freedom, which fundamentally affects the principal nature of the Cooper pair wave function. We investigate the onset of superconductivity in the Rashba-Hubbard model on the triangular lattice, which is symmetry-classified by the associated irreducible representations (irrep) of the hexagonal point group. From an instability analysis by means of the truncated-unity functional renormalization group (TUFRG) we find the $E_2$ irrep to dominate a large fraction of phase space and to lead up to an energetically preferred gapped, chiral superconducting state. The topological phase space classification associated with the superconducting order parameter obtained from TUFRG reveals a fragmentation of the $E_2$ domain into different topological sectors with vastly differring Chern numbers. It hints at a potentially applicable high sensitivity and tunability of chiral superconductors with respect to topological edge modes and phase transitions.
\end{abstract}

\maketitle


{\it Introduction.---}Topological superconductors (TSCs) are amongst the most desired states of topological matter\,\cite{kallin-16rpp054502,sato-17rpp076501}, as their ability to host Majorana zero modes is believed to be key for future implementation of topological quantum computing platforms. This immense technological potential is hindered by the lack of fundamental understanding of TSCs: (i) Theoreticians are still unable to provide recipes or guidance for crystal synthesis or to predict materials hosting TSC phases. (ii) Experimental discoveries of TSC candidate materials often come as a surprise; and often enough after some initial excitement, subsequent studies suggest that trivial superconducting states might be a more likely explanation.
To remedy the former, it is essential that the many-body techniques which predict unconventional superconducting ground states are also capable of shedding light on the topological features of these many-body instabilities.

We are particularly interested in topological superconductivity in the two-dimensional triangular lattice as the simplest case with hexagonal symmetry. Here we can also draw on extensive and noteworthy literature on unconventional superconductivity in the triangular lattice Hubbard model,
motivated by the doped Mott insulator $\kappa$-(BEDT-TTF)$_2$Cu$_2$(CN)$_3$\,\cite{vojta1999}
as well as water-intercalated sodium cobaltates Na$_x$CoO$_2$$\cdot y$H$_2$O\,\cite{honerkamp2003,ikeda2004,zhou2008,chen2013a}. Another important class of materials are the $\sqrt{3}\times\sqrt{3}$ reconstructed adatom lattices on semiconductor substrates such as Sn/Si(111), Pb/Si(111) and Sn/SiC(0001). These  are believed to be good realizations of one-orbital Hubbard models, and their observed ground states range from spin and charge density waves to chiral superconductivity\,\cite{glass2015,badrtdinov-16prb224418,ming2017,cao2018,nakamura2018,tresca2018,adler2019,wolf2022c,machida2022,biderang2022,ming-23np500}. Most recently, the advent of twisted two-dimensional  materials led to a renewed interest in unconventional superconductivity on the triangular lattice\,\cite{classen2019,venderley2019,gneist2022a,scherer2022}.

Inversion-symmetry breaking is ubiquitous in materials, e.g. due to crystalline absence of an inversion center, as in so-called non-centrosymmetric materials, or heavy atom superlattices, heterostructures and surface systems.
Nevertheless, the effect of inversion symmetry breaking, usually manifested as Rashba spin-orbit coupling (RSOC)\,\cite{gorkov2001,vafek-11prb172501,fischer2023} on the correlated and superconducting phase diagrams of such materials and corresponding models remains relatively unexplored.

There have been a few studies on the paradigmatic square lattice case\,\cite{greco-18prl177002,wolf2022c,beyer2023}.
As we will see later in the group-theoretical discussion, the possible mixed states in the triangular lattice case are significantly more interesting.
Notable exceptions where spin-orbit effects were considered in interacting triangular lattice systems are works on $\sqrt{3}\times\sqrt{3}$-Pb/Si(111)\,\cite{badrtdinov-16prb224418,tresca2018,machida2022,biderang2022a}, on $\sqrt{3}\times\sqrt{3}$-Sn/SiC(0001)\,\cite{glass2015} as well as on twisted bilayer systems \ce{WSe2}\,\cite{klebl2023} and \ce{PtSe2}\,\cite{klebl2022a}.

In this Letter, we close this gap in the literature by presenting a paradigmatic case study for the triangular-lattice Rashba--Hubbard model (TLRHM).
Despite being the most basic model, it turns out to feature a surprisingly rich phase diagram.

We employ the truncated united extension of the functional renormalization group (TUFRG). TUFRG allows the investigation of competing many-body instabilities in a way that is unbiased to any one particular instability, treating particle-particle and particle-hole instabilities on equal footing. (For more details, see Supplementary Material\,\cite{sm}, and Refs.\,\cite{metzner2012, platt2013, lichtenstein2017, husemann2009, wang2012, beyer2022} contained therein.) This method is particularly suited to study the effect of RSOC on the interacting phase diagram, as the truncated unity approach retains the details of the involved lattice harmonics of the superconducting pairing.
That allows us to easily extract the degree of singlet-triplet mixing of superconducting states (expected when inversion symmetry is broken\,\cite{gorkov2001}). By feeding the TUFRG output into a Bogoliubov--de Gennes (BdG) formalism, we derive the topological properties of the unconventional superconducting phases.


{\it Model.---}In the following, we investigate the TLRHM defined as
\begin{equation}\label{eq:ham}
    H=
        -\!\!\sum_{ \langle ij \rangle, ss'}
         \!\left[ t \delta_{ss'}  + \alpha \left( {\bvec{\sigma}}
                \times \bvec{\rho}_{ij} \right)_{ss'}^z \right]
          c^\dag_{i s} c^{\phantom{\dag}}_{j s'} +  U\sum_i n_{i\up}n_{i\dw}.
\end{equation}
The sum $\braket{ij}$ runs over nearest neighbor sites, $s$, $s'$ are spin indices, $t$ is the hopping amplitude and $\alpha$ quantifies RSOC. The bond vector is given by $\boldsymbol{\rho}_{ij} = \bvec r_i - \bvec r_j$, ${\bvec\sigma}$ is the vector of Pauli matrices, and $n_{i s} = \ann{i s}\cre{i s}$ is the number operator.

The band structure of \eqref{eq:ham} is readily derived from the Bloch matrix
\begin{equation*}
    h_0^{ss'}(\bvec k) =
    \begin{pmatrix}
        \eps_0(\bvec k) &
        \alpha \left[
            \frac{\partial \eps_0}{\partial k_y} +
            i\frac{\partial \eps_0}{\partial k_x}
            \right] \\
        \alpha \left[
            \frac{\partial \eps_0}{\partial k_y} -
            i\frac{\partial \eps_0}{\partial k_x}
            \right]
        & \eps_0(\bvec k)
    \end{pmatrix}_{ss'}
\end{equation*}
with $\eps_0(\bvec k) = - 2 t[\cos(\bvec a_1 \bvec k)+\cos(\bvec a_2 \bvec k)+\cos((\bvec a_2- \bvec a_1) \bvec k)]$; $\bvec a_1$ and $\bvec a_2$ are the primitive lattice vectors.

We analyse the leading many-body instabilities of the Hubbard model through the use of TUFRG\cite{metzner2012, platt2013}. FRG interpolates between the bare Hubbard interaction and a low-energy effective two-particle interaction vertex, by means of iteratively integrating a flow equation. In the TU formalism, the interaction vertices are expressed in terms of the lattice harmonics\,\cite{lichtenstein2017,husemann2009,wang2012}, which gives the advantage of explicit formulations of leading instabilites. We can then track changes in these tight-binding parameters with changes in normal state parameters across our phase diagrams.

Superconducting instabilities on the triangular lattice can generically be written as\,\cite{sigrist1987}
\begin{equation}
    \hat{\Delta} (\bvec k) = [\Psi(\bvec k) \hat{\mathbb{1}} + \bvec d(\bvec k)
        \cdot {\hat{\bvec{\sigma}}}] i \hat{\sigma}_y \, ,
\end{equation}
where the Pauli matrices $(\hat{\mathbb{1}}, \hat{\sigma}_i)$ act on the spin
subspace.

In the absence of RSOC, spin rotational symmetry admits spin-singlet $\Psi$ and triplet $\bvec d$ states, and
all possible superconducting instabilities are then characterized according to the irreps of the point group \sym D {6h}.

For finite RSOC, the point group symmetry is reduced to \sym C {6v} due to broken inversion symmetry. The spins are also ``frozen'' to the lattice and must rotate with it, \ie the system is now spin-orbit coupled. Thus the superconducting states can no longer be characterized by the spatial irreps but only by {\it total irreps}, i.e., space and spin combined. A complete discussion of group theory is presented in the Supplementary Material\,\cite{sm} (and Ref.\,\cite{sigrist1991} contained therein), including derivation and the complete form of the basis functions; here we wish to emphasize that amongst the total irreps there are three different possibilities for the two-dimensional irrep $E_2$, which are permitted to mix among possible tight-binding superconducting instabilities.
These three states are: the standard $d$-wave spin-singlet $\{\Gamma_{d_{x^2-y^2}}, \Gamma_{d_{xy}}\}$; and two spin-triplet states $\{(\Gamma_f,0,0), (0,\Gamma_f,0)\}$ and $\{(\Gamma_{p_x},-\Gamma_{p_y},0), (\Gamma_{p_y}, \Gamma_{p_x},0)\}$, where the 3-vector corresponds to $\bvec d = (d_x, d_y, d_z)$\,.
We also point out that the superconducting instability that is parallel to the Rashba vector $\bs{g} = ( \partial \varepsilon_0 / \partial k_y, - \partial \varepsilon_0 / \partial k_x, 0)$ is $\bs{d} = ( \Gamma_{p_y}, -\Gamma_{p_x}, 0)$, which is a basis function of the $B_2$ irrep\,\cite{frigeri-04prl097001,fischer2023}.

\begin{figure}[t!]
 \centering
 \includegraphics[width=\columnwidth]{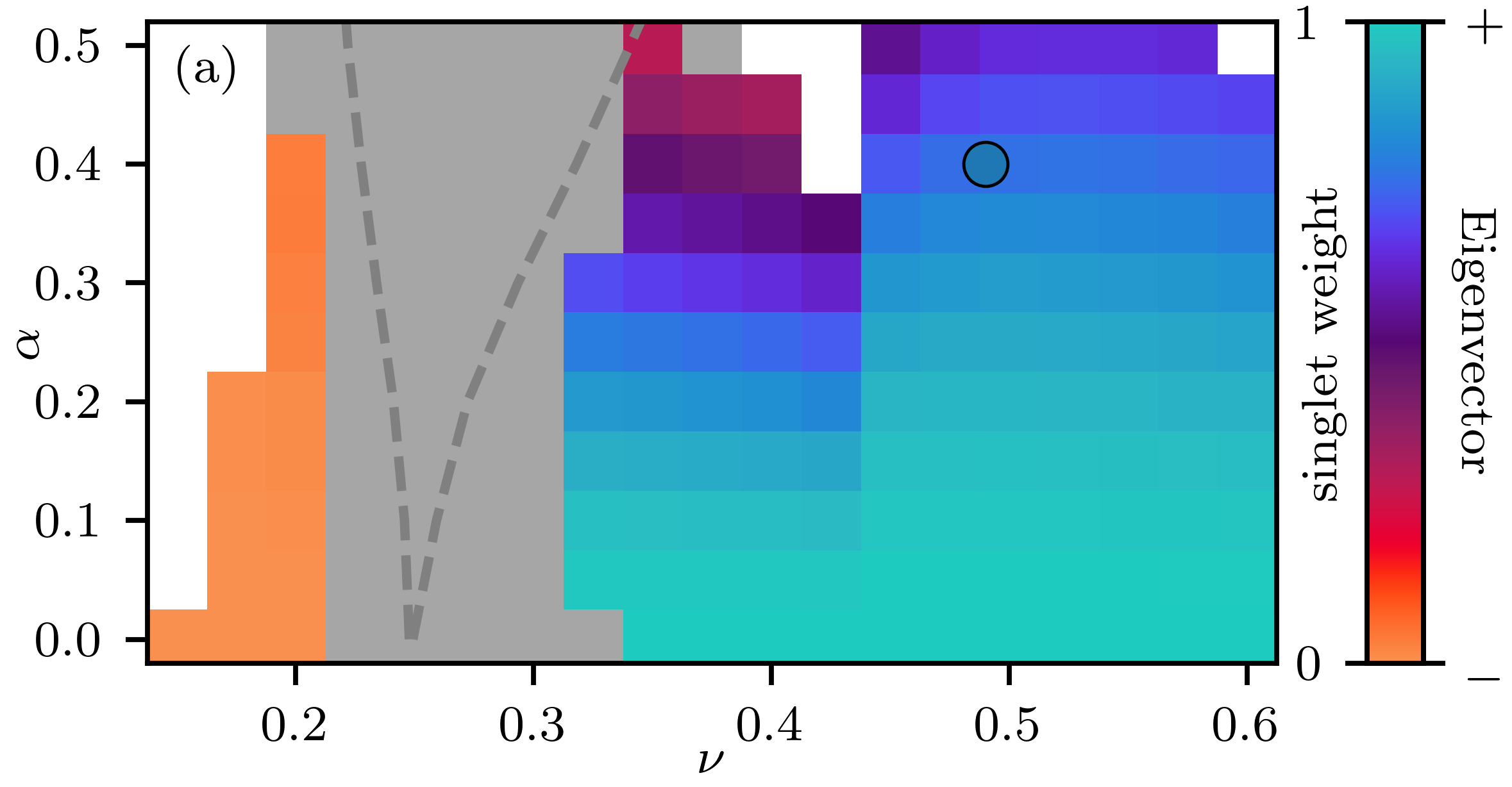}
 \includegraphics[width=\columnwidth]{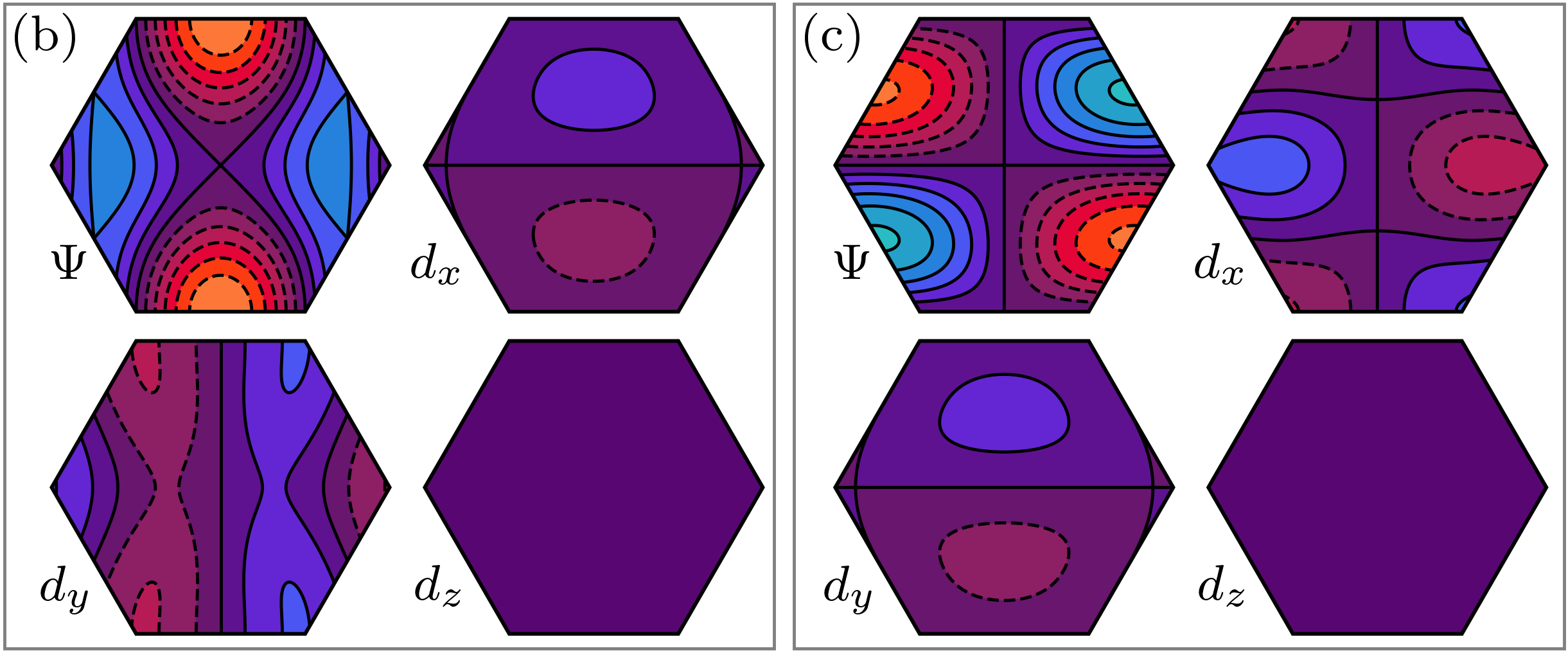}
 \caption{
 (a) Phase diagram with RSOC ($\alpha$) vs.\ filling ($\nu$). Colored areas correspond to superconductivity, grey to magnetic order, and white to metal/Fermi liquid. Color corresponds to singlet-triplet weight of the divergent superconducting instability (see main text). The dashed line indicates van-Hove filling. (b, c) Two degenerate superconducting states corresponding to the parameters as indicated by the circle in (a) $\nu = 0.5, ~ \alpha=0.4$, realizing a superposition of the three possible $E_2$ superconducting states for Cooper-pairs across the nearest-neighbor bonds. The imaginary component of the two superconducting states is zero.}
 \label{fig:phasedia}
 \end{figure}

{\it Results I: competing many-body phases.---}In the following, we present our TUFRG results for $U=8t$ and first focus on different types of instabilities. The phase diagram Fig.\,\ref{fig:phasedia}\,(a) reveals for zero RSOC ($\alpha=0$) and filling $\nu \geq 0.34$ 
an extended superconducting phase with $E_{2g}$ irrep of the symmetry group D$_{6h}$ (spin-singlet), adjacent to a magnetically ordered phase at lower fillings.
There is also a spin-triplet superconducting phase transforming under the $B_{1u}$ irrep for fillings $\nu < 0.22$, below the van Hove singularity.
The extended $E_{2g}$ phase is not unexpected, as this is the preferred type of unconventional superconductivity on hexagonal lattices with a spherical Fermi surface.

The two-dimensional $E_{2g}$ irrep allows for arbitrary complex superpositions of the two degenerate superconducting states that constitute the diverging instability.
Free energy calculations reveal that the equal weight complex superposition ``$d+id$\,'' has the lowest energy (see Supplementary Material (and Ref.\,\cite{klebl2023}) for explicit calculations), due to the largest condensation energy (as previously found in Refs.\,\onlinecite{sigrist1987, sigrist1991, kuznetsova2005} or often assumed without explicit calculation).
We stress that the $d+id$ SC is a topological state with chiral edge states and finite Chern number, as explicitly confirmed below.
At first view, the $\alpha=0$ results are compatible with previous work\,\cite{honerkamp2003, klebl2023, classen2019}; our analysis below reveals, however, that part of it is an {\it extended} chiral $d$-wave state.
Outside of the shown fillings $\nu$, TUFRG does not diverge, signaling stability of the Fermi liquid phase.

Increasing the RSOC $\alpha$ does not change much on the interplay of superconducting and magnetic phases in the considered range $0\leq\alpha<0.5$ (see Fig.\,\ref{fig:phasedia}\,a). We do find commensurate and incommensurate spin-density waves with both ferro- and antiferromagnetic ordering tendencies, but leave the detailed analysis of the magnetically ordered phase for future work.

{\it {Results II: singlet-triplet mixing}.---} In line with previous work in Rashba systems\,\cite{vafek-11prb172501,wolf2020,beyer2023},
we find the fraction of singlet-triplet mixing to increase with increasing RSOC, so long as the SC phase is spin singlet at $\alpha = 0$.
One of the major advantages of TUFRG is to quantitatively determine this singlet-triplet mixing (see Supplementary Material\,\cite{sm} for details).
In particular, for $0.34 \leq \nu \leq 0.44$, finite $\alpha$ and thus inversion symmetry breaking, moves the chiral $d+id$ SC phase at $\alpha = 0$ to the $E_2$ irrep of $C_{6v}$, allowing an admixture of this $d+id$ spin singlet state and the two triplet $E_2$ states discussed above.
We observe sizeable triplet contributions in the otherwise singlet-dominated phase, and for larger values of RSOC, such as $\alpha=0.5$, singlet and triplet contributions are of similar weight.
An example of this mixed singlet-triplet order parameter is shown in Fig.\,\ref{fig:phasedia}\,(b,c). The superconducting order parameter was calculated for half-filling $\nu = 0.5$ and $\alpha = 0.5$, and transforms under the two-dimensional $E_2$ irrep. In this particular example, the $d$-wave singlet component $\Psi$ is the largest, with a small mixture of the two triplet states. This mixture distinguishes all order parameters for $\alpha > 0$ from the exclusively spin-singlet chiral $d+id$. The breakdown of this state into allowed form factors is discussed further in the Supplementary Material\,\cite{sm}, in particular see Fig.\,S2.

In contrast to the superconducting phase for fillings above the vH singularity, the low-filling superconducting phase remains fully spin-triplet regardless of $\alpha$. However, also here we observe a mixture of SC order parameters which is not possible for $\alpha=0$: the two  triplet $E_2$ superconducting states mix, forming a triplet-triplet mixture, yielding the extended triplet phase.
%
%
RSOC does, however, change the nature of the triplet state. At $\alpha=0$ we find $B_{1u}$ irrep (gapless), and for any finite $\alpha$ the $E_2$ irrep containing $f$-wave and $p$-wave pairings. Roughly speaking, the $p$-wave contribution, responsible for the finite gap, increases with $\alpha$. At $\alpha\sim 0.4t$ this $E_2$ phase is dominated by $p$-wave pairing. We stress that there is hardly any method which can resolve spin-triplet or triplet-triplet mixing as easily as TUFRG can. To determine whether the mixed-in spin-triplet components trigger any additional topological features we need to push the analysis further.

\begin{figure}[t!]
\centering
\includegraphics[width=\columnwidth]{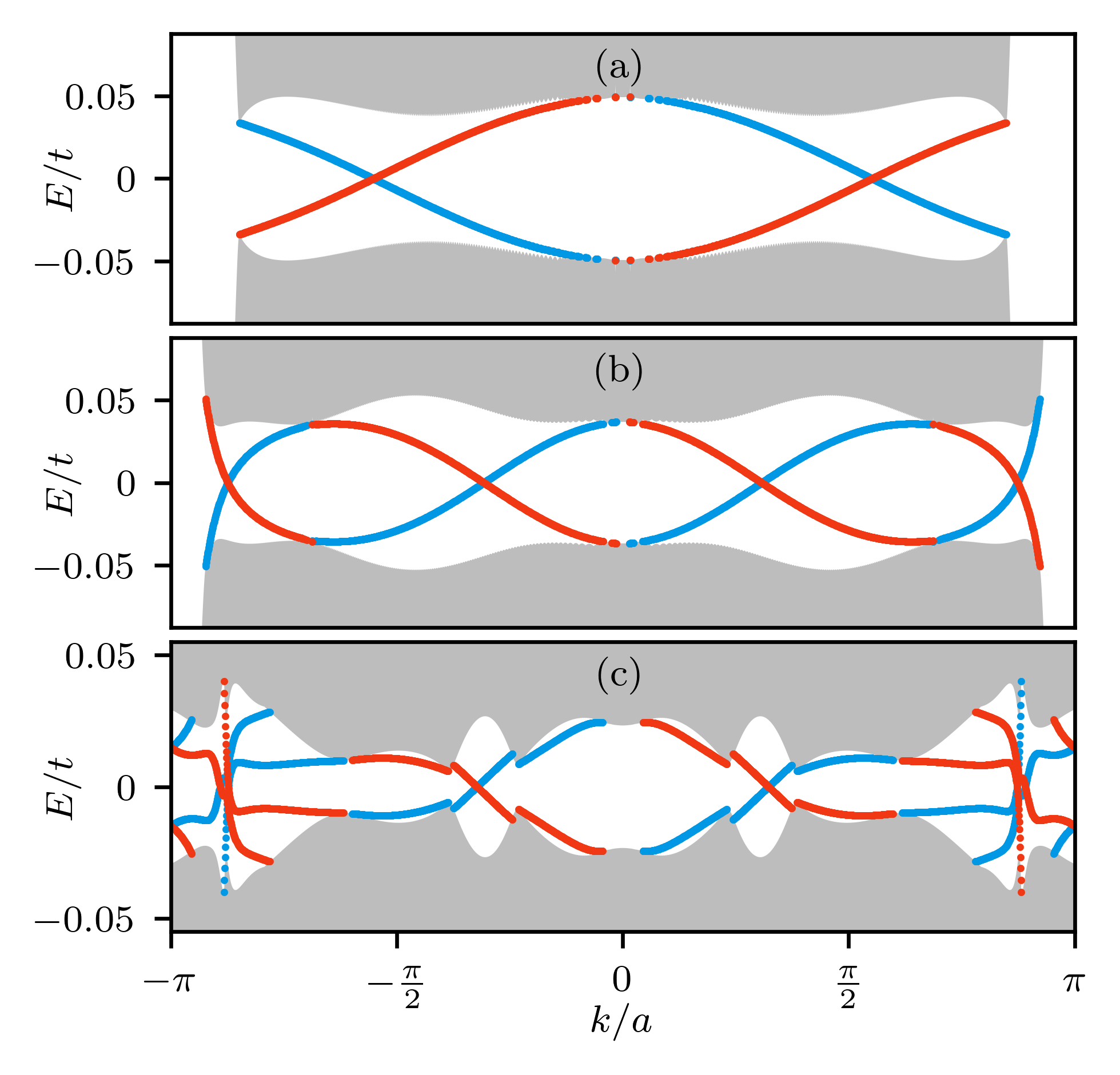}
\caption{Ribbon spectra from TUFRG and BdG methods.
(a) $\alpha=0$, $\nu=0.5$ resulting in $C=4$, as evidenced by two doubly degenerate chiral edge modes {\it per edge}.
(b) $\alpha=0$, $\nu=0.4$ resulting in $C=-8$, as evidenced by four doubly degenerate chiral edge modes {\it per edge}.
(c) $\alpha=0.35$, $\nu=0.4$ resulting in $C=-6$, as evidenced by six chiral edge modes {\it per edge}.
All spectra calculated for a SC amplitude of $|\Delta| = 0.05$, on a ribbon of 900 atoms, and a $k$-momentum resolution of 3000 points.
}
\label{fig:ribbon}
\end{figure}

{\it {Results III: BdG analysis, Chern numbers and ribbon spectra}.---}TUFRG provides us with details about the pairing symmetry and relative strength for shells with different distance (\eg on-site pairing vs.\ nearest-neighbor pairing), thanks to its truncated unity extension. In the following, we directly make use of this information and feed it into a BdG Hamiltonian (see Supplementary Material\,\cite{sm}) where all other bandstructure parameters are kept identical to the TUFRG input. Since the superconducting states are all $E_2$ and doubly degenerate, we calculate the free energy to determine the stable ground state configuration\,\cite{platt2013, klebl2022a, classen2019}. Knowledge of the BdG Hamiltonian and its eigenvectors allows us to compute Chern numbers of the corresponding gapped superconducting phases and compute ribbon (and real space) spectra to reveal topological edge states. For instance, by choosing $\alpha=0$ and $\nu=0.5$ we recover the earlier result\,\cite{crepieux2023} that
the chiral $d+id$ superconductor ($E_2$ irrep) possesses a Chern number $C=4$. The corresponding ribbon spectra is shown in Fig.\,\ref{fig:ribbon}\,(a), which at $\alpha=0$ manifests as only two distinguishable edge states due to spin degeneracy\cite{black-schaffer2012a, scherer2022}. When a small amount of RSOC is introduced, \ie $\alpha > 0$, the spin degeneracy is lifted and four distinguishable chiral edge states are observable.

\begin{figure}[t!]
\centering
\includegraphics[width=\columnwidth]{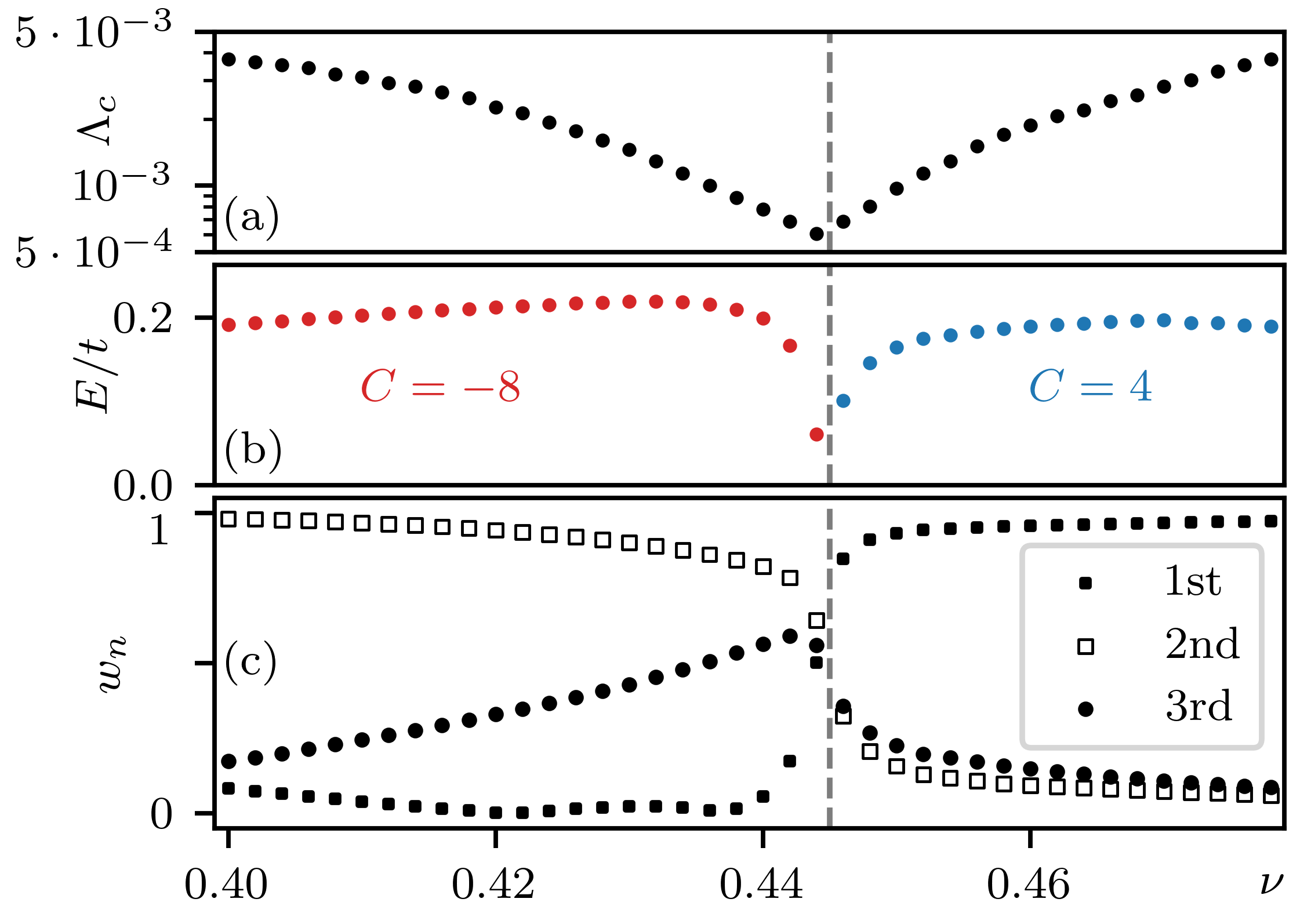}
\caption{Topological transition in filling at \texorpdfstring{$\alpha = 0$}{zero alpha}. (a) Critical TUFRG scale $\Lambda_c$ as a function of filling $\nu$. (b) BdG energy $E$ as a function of filling $\nu$, Chern numbers color-coded relative to Fig.\,\ref{fig:topo-phasedia}. (c) Relative weights in the TUFRG superconducting instability for the $n$-th nearest neighbor pairing lengths as a function of filling. In all plots, the topological transition at filling $\nu = 0.445$ is shown by the dashed grey line.
}
\label{fig:n-transition}
\end{figure}

{\it Topological phase transitions.---}By virtue of combining the TUFRG and BdG methods, we analyze the entire superconducting phase space. First, we focus on constant RSOC $\alpha$ and vary the filling $\nu$. At $\nu_c = 0.445$ (irrespective of $\alpha$), we observe a dip in critical scale $\Lambda_c$ (see Fig.\,\ref{fig:n-transition}\,(a)). We note that this feature should also be observable within other FRG approaches. We observe this non-analytical behavior also in other quantities such as the gap energy of the BdG Hamiltonian (Fig.\,\ref{fig:n-transition}\,(b)), as well as a shift in the spectral weight of the Cooper pairing from second- and third-nearest neighbor bonding to primarily nearest-neighbor. We visualize this by calculating the spectral weight on each interaction shell $w_n$ in Fig.\,\ref{fig:n-transition}\,(c) (for details see Supplementary Material\,\cite{sm}). We find that for fillings larger than $\nu_c$ the BdG Hamiltonian 
possesses a Chern number $C=4$, however, for fillings smaller than $\nu_c$ we obtain $C=-8$; see the corresponding ribbon spectra in Fig.\,\ref{fig:ribbon} (a, b), revealing a topological phase transition.

We can understand this topological phase transition by focusing on the purely singlet chiral $d+id$-wave superconducting phases at $\alpha = 0$. In this limit, the gap in the BdG spectrum closes when the Fermi surface intersects vortices in the superconducting pairing $\Gamma_{d_{x^2 - y^2}} + i \, \Gamma_{d_{xy}}$. The location and number of these vortices depends on the precise mix of the three interaction shell pairings. With a longer pairing, higher harmonics are introduced into the superconducting pairing function, which means more vortices are introduced for longer range pairing. These vortices can then be shifted in reciprocal space by changing the admixture of the superconducting pairings, with the total number changing when vortices recombine and are moved in and out the Brillouin zone. That is, the number of vortices can change and thus the topological invariant can change too. Bottomline: even in the absence of RSOC, topological phase transitions are to be expected, as demonstrated above, although we find $E_2$ irreps on both sides of the phase transition. We note that a similar phase transition was observed at $\alpha=0$ in extended Hubbard and $tJ$ models\,\cite{scherer2022, valkov2015, zhou2008}. Our work establishes the plain-vanilla onsite Hubbard model as the minimal model to find such a phase transition, as well as its stability towards RSOC.

\begin{figure}[t!]
\centering
\includegraphics[width=\columnwidth]{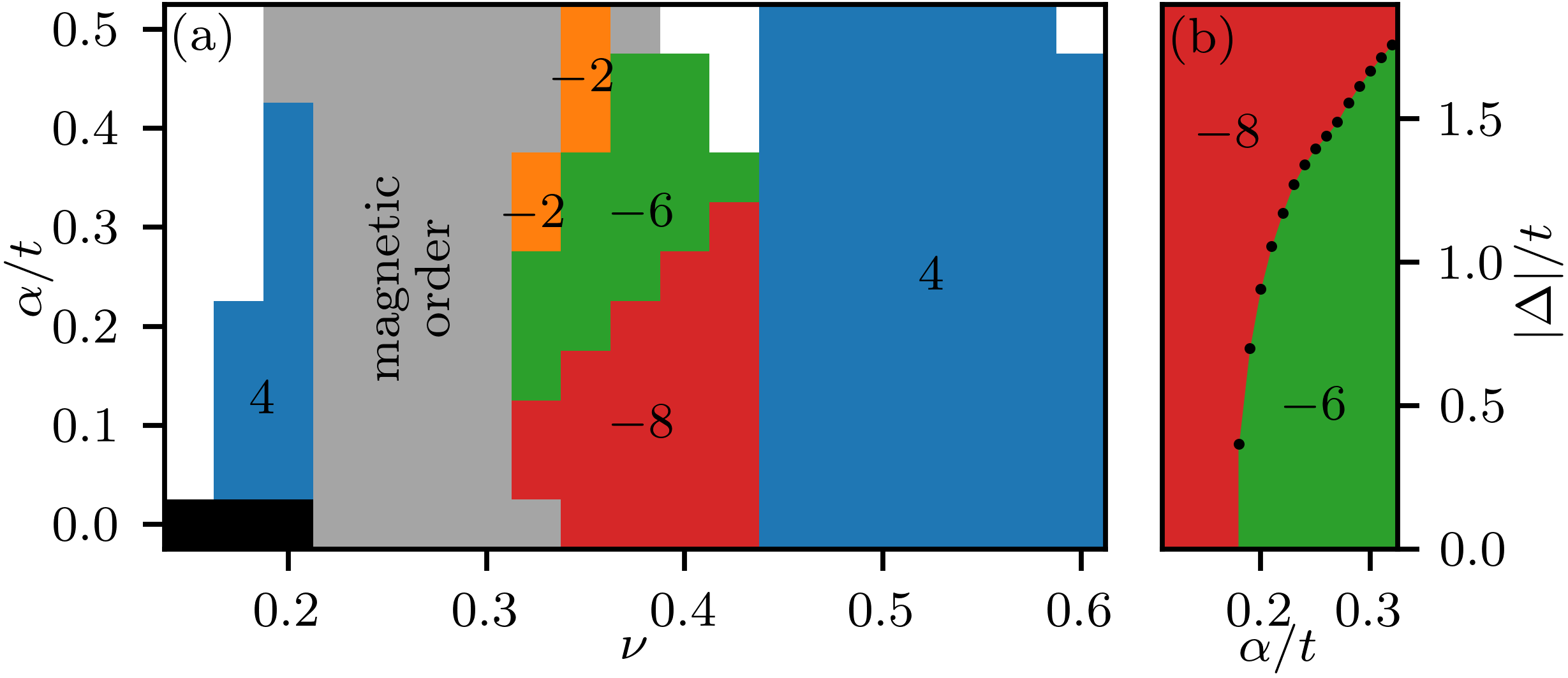}
\caption{(a)  Topological phase diagram: superconducting phases from Fig.\,\ref{fig:phasedia}\,a with their Chern numbers. White regions correspond to the Fermi liquid, grey is magnetic ordering, and black states are gapless. (b) Varying Chern number with increasing RSOC $\alpha$ and superconducting order parameter amplitude $|\Delta|$, for fixed filling $\nu = 0.4$.
}
\label{fig:topo-phasedia}
\end{figure}

We also encounter topological phase transitions when varying $\alpha$. For instance, for fixed $\nu=0.4$, we previously found $C=-8$ at $\alpha=0$ (see a corresponding ribbon spectra in Fig.\,\ref{fig:ribbon}\,(b)) and for larger $\alpha=0.35$ we find $C=-6$, as corroborated by Fig.\,\ref{fig:ribbon}\,(c). The entire analysis of Chern number and edge states from ribbon spectra culminates in the topological phase diagram Fig.\,\ref{fig:topo-phasedia}\,(a).

Clear critical signatures as those shown in Fig.\,\ref{fig:n-transition}\,(a) are absent, however, the BdG energy gap clearly reveals the topological phase transition when changing $\alpha$.
This difference can be traced to the nature of the BdG gap closing. Unlike the purely singlet case, the characteristic polynomial of the BdG matrix has some constant terms, which mean that the BdG energies can have zeros and the BdG gap can close where the superconducting amplitude $|\Delta|$ is not zero; Fig.\,\ref{fig:topo-phasedia}\,(b) plots the relationship between $\alpha$ and $|\Delta|$.  If we assume that $|\Delta|/t < 0.1$ (which is reasonable as we follow the educated guess $|\Delta| \propto \Lambda_c$) then this fixes $\alpha_c = 0.17$ to two significant figures regardless how small $|\Delta|$ will be. This also justifies omitting an additional self-consistent treatment of the magnitude of $|\Delta|$.
We then compute topological invariants and derive ribbon spectra resulting in Fig.\,\ref{fig:topo-phasedia}\,(a).

{\it Outlook.---}There are many examples of studying unconventional superconductivity as the groundstate of correlated electron systems in the literature. In particular, for hexagonal lattices with a circular Fermi surface, the $E_2$ irrep is the most likely candidate. While it is well-known that such a state usually corresponds to the chiral $d+id$ superconductor, here we show that the phenomenology is much richer, in particular, in the presence of spin-orbit coupling. We find Chern numbers $C= 4, -2, -6, -8$ within an otherwise {\it homogeneous} $E_2$ irrep phase. Only the employed TUFRG formalism allows to reveal these topological features and the associated topological phase transition within that phase. In the light of ongoing experimental activities with the prospect of unconventional superconductivity in triangular lattice materials such as B-doped Sn/Si(111) and related compounds, we hope that our work will spark further activity for these systems since different topological states and possibly even topological phase transitions might be experimentally observable.

\begin{acknowledgments}
The authors thank L.\ Classen, M.\ D\"urrnagel, A.\ Fischer, L.\ Klebl, J.\ B.\ Profe, A.\ Sanders, A.\ Schnyder, T.\ Schwemmer and C.\ Timm for helpful discussions.
This work was funded by the Australian Research Council (ARC) through Grant No.\ DP200101118, by the German Research Foundation (DFG) through RTG 1995 and the Priority Program SPP 2244 \enquote{2DMP} and
by the W\"urzburg--Dresden Cluster of Excellence on Complexity and Topology in Quantum Matter – ct.qmat (EXC 2147, project id 390858490). This research was supported in part by the National Science Foundation under Grant No.\ NSF PHY-1748958.
The authors gratefully acknowledge the scientific support and HPC resources provided by the Erlangen National High Performance Computing Center (NHR@FAU) of the Friedrich-Alexander-Universität Erlangen-Nürnberg (FAU) under the NHR project \enquote{k102de--FRG}. NHR funding is provided by federal and Bavarian state authorities. NHR@FAU hardware is partially funded by the DFG – 440719683. The authors further acknowledge the resources from the National Computational Infrastructure (NCI Australia), an NCRIS enabled capability supported by the Australian Government.
\end{acknowledgments}

\bibliography{refs_TLRHM}

\end{document}


\title{Supplementary Material:\texorpdfstring{\\}{}
Chern number landscape of spin-orbit coupled chiral superconductors}
\author{Matthew Bunney}
\thanks{These authors contributed equally.}
\affiliation{School of Physics, University of Melbourne, Parkville,
    VIC 3010, Australia}
\affiliation{Institute for Theoretical Solid State Physics,
    RWTH Aachen University, 52062 Aachen, Germany}

\author{Jacob Beyer}
\thanks{These authors contributed equally.}
\affiliation{School of Physics, University of Melbourne, Parkville,
    VIC 3010, Australia}
\affiliation{Institute for Theoretical Solid State Physics,
    RWTH Aachen University, 52062 Aachen, Germany}
\affiliation{JARA Fundamentals of Future Information Technology,
    52062 Aachen, Germany}
\affiliation{Institute for Theoretical Physics, University of Würzburg,
    Am Hubland, 97074 Würzburg, Germany}

\author{Ronny Thomale}
\affiliation{Institute for Theoretical Physics, University of Würzburg,
    Am Hubland, 97074 Würzburg, Germany}

\author{Carsten Honerkamp}
\affiliation{Institute for Theoretical Solid State Physics,
    RWTH Aachen University, 52062 Aachen, Germany}
\affiliation{JARA Fundamentals of Future Information Technology,
    52062 Aachen, Germany}

\author{Stephan Rachel}
\affiliation{School of Physics, University of Melbourne, Parkville,
    VIC 3010, Australia}

\date{\today}

\maketitle

\section{TUFRG method}

\begin{figure}
    \centering
    \includegraphics[width=0.45\textwidth]{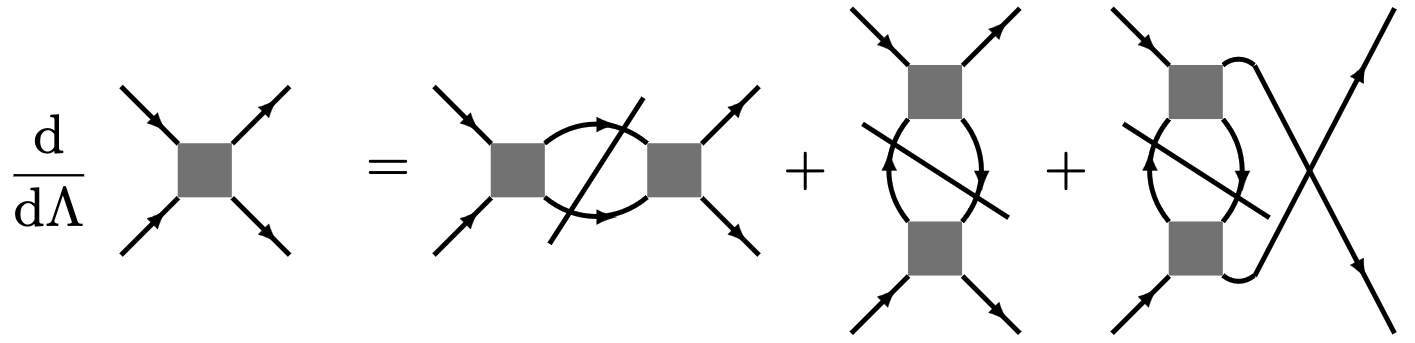}
    \caption{
        \captiontitle{Diagrammatic representation of the FRG flow equation.}
        We show the non-\su{} flow equation up to the two-particle effective
        interaction, up to $U^2$.
        The nodes in the diagrams represent the effective interaction at scale
        $V^\Lambda$, while the connecting lines are loop derivatives w.r.t. the
        scale
        $\dot L^{\Lambda} =
        [G^{\Lambda}\frac \dd {\dd \Lambda}G^{\Lambda}
            - \frac \dd {\dd \Lambda}G^{\Lambda}G^{\Lambda}]$ with the permutation
        of indices implied.}
    \label{fig:diags}
\end{figure}

The functional renormalization group is a powerful tool for calculating the leading instabilities of interacting Fermi liquid systems without bias between the competing interaction channels.
In our numerical calculations, this is practically achieved by solving a flow equation for an effective four-particle interaction vertex $V^\Lambda$, regularized by the scale or flow parameter $\Lambda$.
The scale $\Lambda$ is artificially inserted into the flow equation via the Green’s function $G^\Lambda = \Theta^\Lambda G$, where the ``$\Omega$-cutoff'' $\Theta^\Lambda = \frac{\Lambda^2}{\Lambda^2 + \Omega^2}$ is our regularization function. A diagrammatic representation of the flow equation is contained in Fig.\,\ref{fig:diags}; see also these comprehensive review papers\,\cite{metzner2012,platt2013}.
We use a static formulation of FRG, neglecting self-energy and higher-order interaction vertex corrections.

We solve the flow equation from high scale $\Lambda \rightarrow \infty$, corresponding to our bare interaction vertex, \ie the onsite Hubbard interaction in Eq.\,(1), iteratively down towards $\Lambda = 0$. However, often during the flow, one of the physical modes of the vertex diverges at a critical scale $\Lambda_c$, signifying the onset of a phase transition into a spontaneously symmetry-broken (SSB) state. This SSB instability can be extracted from the effective vertex at critical scale $V^{\Lambda_c}$ by diagonalizing in the correct physical channel. For example, a divergence in the particle-particle effective vertex $V^{\Lambda_c}_P$ with transfer momentum $\bvec q=0$ would yield a superconducting instability (with eigenvalue $\lambda$):
\begin{equation} \label{eqn:frg_sc}
    \Delta_{s_1 s_2} (\bvec{k}) = \frac{1}{\lambda} \sum_{\bvec{k}^\prime s_3 s_4}
    V^{\Lambda_c}_{P, s_1 s_2 s_3 s_4} (\bvec{q} = 0, \bvec{k}, \bvec{k}^\prime) \Delta_{s_3 s_4} (\bvec{k}^\prime) \, ,
\end{equation}
where $\Delta_{ss'}(\bvec{k})$ are the components of the superconducting order parameter $\hat\Delta(\bvec{k})$.

The weak-coupling regime where FRG is considered to be a controlled approach is typically at $U/W \sim 1/2$, and around $U/W \approx 1$ one typically might be leaving this regime. Here we performed at $U=8t$ which corresponds to $U/W=1$ at $\alpha=0$ up to $U/W\approx 0.92$ at $\alpha=0.5$. That is, one might argue that we are no longer in a weak-coupling regime where FRG is truly controlled. Yet, the reason for choosing these $U$-values is that this generates pairing tendencies at numerically resolvable scale and thus provides a qualitative overview over the main perturbative tendencies of the system. While we expect that quantitative properties would experience corrections if less approximate approaches than our FRG were used, we believe that the ordering tendencies described here remain present. Comparing the data at $U=8t$ with that for smaller $U$ for a number of cases furthermore indicates that the behavior described here is robust for a larger range of interaction strengths. Moreover, by choosing $U=8t$ we avoid the chiral spin liquid phase at half filling for $U>8.5t$\,\cite{szasz-20prx021042}.

We employ the truncated unity extension of the functional renormalization group\,\cite{lichtenstein2017,husemann2009,wang2012}. The core idea is that we series-expand the relative momenta dependence of our vertex into form factors, which are most conveniently chosen in terms of Bravais lattice vectors $\bvec f$,
\begin{equation}
    V_{ff'} (\bvec q) = \sum_{\bvec{k}, \bvec{k}'} e^{ - i \bvec{f} \cdot \bvec{k}} e^{ i \bvec{f'} \cdot \bvec{k'}} V (\bvec q, \bvec k, \bvec k') \, ,
\end{equation}
where we have neglected the spin dependence of the vertex. If we consider the entire lattice, this is a unitary transformation, however we truncate the full series expansion to only short-range form factors, which we argue captures all relevant physical results, \ie
\begin{equation}
     V (\bvec q, \bvec k, \bvec k')  \approx \sum_{|\bvec{f}|, |\bvec{f}'| < F} e^{i \bvec{f} \cdot \bvec{k}} e^{ - i \bvec{f'} \cdot \bvec{k'}} V_{ff'} (\bvec q) \, ,
\end{equation}
for some length cutoff $F$. We establish convergence by confirming that the results of the FRG calculation do not change with an increase in the truncation range.
This truncation of the unitary expansion of the vertex reduces computational complexity significantly, while retaining the momentum conservation of the vertex required to accurately capture spin-momentum locking.
The TUFRG code used for the simulations in this manuscript was benchmarked against other codes as code \#2 in Ref.\,\onlinecite{beyer2022}. This reference also provides information about the structure of the code base.

The TU formulation also adds the advantage of handling the vertex and superconducting instabilities in terms of the lattice harmonics, allowing us to directly access the functional forms of the superconducting instabilities, which we do in Sec.\,\ref{app:harmonics}.
The form-factor and momentum space representations of the superconducting instability are related,
\begin{equation}
    \hat{\Delta}_{f} = \sum_{k} e^{- i \bvec f \cdot \bvec k} \hat{\Delta} (\bvec k) \, ,
\end{equation}
and similarly to the momentum space representations, the form-factor SC instability can be decomposed into spin-singlet $\Psi_f$ and spin-triplet $\bvec{d}_f$ components
\begin{equation}
    \hat{\Delta}_f = [\Psi_f \hat{\mathbb{1}} + \bvec d_f \cdot {\hat{\bvec{\sigma}}}] i \hat{\sigma}_y \, .
\end{equation}
Singlet weight of TU SC instability, as plotted in Fig.\,1, can then be calculated as
\begin{equation}
    w_{\text{sing.}} = \frac{1 }{|\Delta|^2} \sum_f | \Psi_f |^2 \, ,
\end{equation}
where we have normalized by the amplitude of the superconducting instability squared:
\begin{equation}
    |\Delta| = \sqrt{ \sum_f \frac{1}{2} \text{Tr} [\hat{\Delta}^\dagger_f \hat{\Delta}^{\phantom{\dagger}}_f]} \, .
\end{equation}
The form-factor representation also allows us to calculate the spectral weight of the TU SC instability on the $n$th interaction shell, as shown in Fig.\,(3)\,(c):
\begin{equation}
    w_{n} = \frac{1}{2 |\Delta|^2} \sum_{f \in S_n } \text{Tr} [\hat{\Delta}^\dagger_f \hat{\Delta}^{\phantom{\dagger}}_f] \, ,
\end{equation}
where $S_n = \{f : |\bvec f| = n \}$.

\section{Group theoretical considerations}

\subsection{Symmetry group of the Hamiltonian}

Without Rashba spin-orbit coupling, the symmetry of the superconducting Hamiltonian is given by:
\begin{equation}
    \sym D {6h} \otimes SU(2) \otimes U(1) \otimes \mathbb{Z}_2
\end{equation}
which are the space group \sym D {6h}, the $SU(2)$ rotations of the spins, $U(1)$ gauge symmetry and a $\mathbb{Z}_2$ time reversal symmetry.

The space group \sym D {6h} consists of the point group -- which is our main concern -- together with translations. The point group has 24 symmetries, and is generated by the rotations $C_6$, $C^{\prime}_2$ and inversion $\mathcal{I}$.
The Hamiltonian (1) with no RSOC is also equivalent under any $SU(2)$ rotation of the spins
\begin{equation}
    c_{k s} \rightarrow U_{s s'} (\theta, \hat{\bvec{n}}) \, c_{k s'} \, ,
\end{equation}
where
\begin{align}
    U_{s s'} (\theta, \hat{\bvec{n}}) &= \left( \exp \left[ - \frac{i \theta (\bvec{\sigma} \cdot \hat{\bvec{n}} ) }{2} \right] \right)_{ss'} \, ,
    \label{eqn:euler_spin}
\end{align}
which is the Euler rotation by angle $\theta$ around principal axis $\hat{\bvec{n}}$.

Introducing the Rashba term affects the symmetry of the Hamiltonian in two ways. First, the Rashba term breaks inversion symmetry, reducing the space group from \sym D {6h} to \sym C {6v}.
Conventionally, \sym C {6v} is generated by the rotation $C_6$ and a mirror plane $\sigma_1$, and realizes all the symmetries of a hexagon embedded in two spatial dimensions.

The second way that the Rashba term affects the symmetry is by coupling the spin rotations to those of the lattice. The freedom to rotate each component independently is lost, as the cross product ${\bvec{\sigma}} \times \bvec{\rho}_{ij}$ in the Rashba term only remains invariant when the spins and lattice are rotated/reflected simultaneously. This term is only invariant under those symmetry operations in the group with the lower symmetry, \ie \sym C {6v}.

The complete symmetry of the Hamiltonian with RSOC is then:
\begin{equation}
    \text{\sym C {6v}} \otimes U(1) \otimes \mathbb{Z}_2
\end{equation}
where \sym C {6v} is the point group--transforming  both the coupled spins and lattice -- together with lattice translations, as well as gauge and time reversal symmetries.

\subsection{Basis functions of irreps}

Since we know that the symmetry group of the superconducting RSOC Hamiltonian is \sym C {6v}, we aim to use this to classify our possible superconducting instabilities.
The superconducting gap is the solution to \cref{eqn:frg_sc}, which is in the form of an eigenvalue problem. The eigenvector space decomposes into a basis of the irreps of the underlying symmetry group of the matrix, therefore the possible superconducting instabilities are the irrep basis functions\cite{sigrist1991}.

The particle-particle vertex diverged at $\bvec q = 0$, as it appears in Eq.\,\eqref{eqn:frg_sc}, acts on the space of Cooper pairs $c^\dagger_{\bvec{k} s} c^{\dagger}_{-\bvec{k} s'}$, which is a Hilbert space indexed by the continuous momenta $\bvec k$ and two spins $s, s'$. In the TUFRG formalism, we replace the continuous momenta $\bvec k$ with the discrete form factors $\bvec f$, so the Hilbert space is indexed by $(f, s, s')$.

The irrep basis functions of \sym C {6v} acting on the Hilbert space $(f, s, s')$ can be found by constructing a matrix representation $R(\sym C {6v})$, then constructing projection operators for each irrep $\rho$ by use of the formula:
\begin{equation}
    P^{(\rho)} = \sum_g \frac{d_\rho}{|G|} \chi^* (g) \, R(g) \, ,
\end{equation}
where $\chi(g)$ is the character of group element $g$, $|G|$ is the order of the group (12 for $\sym C {6v}$), and $d_\rho$ is the dimension of the irrep $\rho$.
The eigenvectors of the projection operator $P^{(\rho)}$ with eigenvalue 1 then form a basis for the irrep $\rho$.

The symmetries act on the superconducting state $\Delta_{f s s'}$ as:
\begin{equation}
    \Delta_{f s s'} \rightarrow \sum_{f' r r'} R_{ff'} U_{s r} U_{s' r'} \Delta_{f' r r'} \, ,
\end{equation}
with spatial transformations $R_{ff'}$, which rotate the form factor indices identically to rotating the corresponding real space bonds, and the Euler rotations of the spins $U_{sr}$, shown in Eqn.\,\eqref{eqn:euler_spin}.

In order to extract the basis functions in a useful form, it is helpful to first find the irrep basis functions for the spatial and spin parts separately, then find the total representations from the tensor product of the irreps of each part.

In the case where the spins maintain a distinct full $SU(2)$ rotational symmetry, superconducting states can be classified by the the irrep basis functions from the representation the spatial pairing $R_{ff'}$.
The basis functions $\Gamma_\ell$ are calculated for each interaction shell. The basis functions for interaction shells with six elements (such as nearest as well as next-nearest neighbor pairings) are listed in the second column of Tab.\,\ref{tab:irrep_basis_fns}.

\begin{table}
\centering
\begin{tabular}{ccc} \toprule
    {$\rho$} & $\Gamma$-Basis & $\sigma$-Basis \\ \hline
    \sym A 1    & \numberedHexagon{$\Gamma_s$}{1}{1}{1}{1}{1}{1}
                & $\Psi = 1$ \\
    \sym A 2    & \textbf{---}  & $\bvec d = (0,0,1)$  \\
    \sym B 1 & \numberedHexagon{$\Gamma_f$}{-1}{1}{-1}{1}{-1}{1} & \textbf{---}  \\
    \sym E 1 &
    \hspace{0.2cm} 
        $\begin{bmatrix}
            \numberedHexagon{$\Gamma_{p_x}$}{1}{-1}{-2}{-1}{1}{2} &
            \numberedHexagon{$\Gamma_{p_y}$}{1}{1}{0}{-1}{-1}{0}
        \end{bmatrix} $
    \hspace{0.2cm} 
        &
        $\begin{bmatrix}
            \bvec d = (1,0,0) \\
            \bvec d = (0,1,0)
        \end{bmatrix}$ \\[3em]
    \sym E 2 &
        $\begin{bmatrix}
            \hspace{0.03cm}
            \numberedHexagon{$\Gamma_{d_{x^2-y^2}}$}{-1}{-1}{2{\phantom -}}{-1}{-1}{2}
            \hspace{0.03cm}
            &
            \hspace{0.03cm}
            \numberedHexagon{$\Gamma_{d_{xy}}$}{1}{-1}{0}{1}{-1}{0}
            \hspace{0.03cm}
        \end{bmatrix}$
        & \textbf{---} \\[3em] \toprule
\end{tabular} \\
\vspace{1em}
  \caption{
        \captiontitle{Basis functions of irreducible representations.}
        The first column lists the irreducible representation (irrep) $\rho$.
        The second column is the basis function in real space, for example, as
        the bond pairing on nearest neighbor bonds. The third column is the
        two-spin basis function, in terms of the typical superconducting
        pseudo-vector formulation $(\Psi, \bvec d)$.}
    \label{tab:irrep_basis_fns}
\end{table}

\begin{table}
\centering
\begin{tabular}{cccc} \toprule
   \diagbox{$\rho_\Gamma$}{$\rho_\sigma$}
       & \sym A 1 & \sym A 2 & \sym E 1 \\ \hline \\
   \sym A 1
       & \textcolor{first}{\sym A 1}: $\Gamma_s$
       & \textbf{---}
       & \textbf{---} \\[1em]
   \sym E 2
       & \textcolor{first}{\sym E 2} : $\begin{bmatrix} \Gamma_{d_{xy}} \\
           \Gamma_{d_{x^2-y^2}}\end{bmatrix}$
       & \textbf{---}
       & \textbf{---} \\[1em] \hline \\
   \sym B 1
       & \textbf{---}
       & \textcolor{first}{\sym B 2} : $(0, 0, \Gamma_f)$
       & \textcolor{first}{\sym E 2} : $ \begin{bmatrix} (\Gamma_f, 0, 0) \\ (0, \Gamma_f, 0) \end{bmatrix}$ \\[2.5em]
   \sym E 1
       & \textbf{---}
       & \textcolor{first}{\sym E 1} :
           $\begin{bmatrix} (0,0,\Gamma_{p_x}) \\ (0,0,\Gamma_{p_y}) \end{bmatrix}$
       & $\begin{gathered}
           \textcolor{first}{\sym A 1 \oplus \sym B 2 \oplus \sym E 2} : \\
           (\Gamma_{p_x}, \Gamma_{p_y}, 0) \\
           \oplus \\
           (\Gamma_{p_y}, -\Gamma_{p_x}, 0) \\
           \oplus \\
           \begin{bmatrix}
            (\Gamma_{p_x}, -\Gamma_{p_y}, 0) \\
            (\Gamma_{p_y}, \Gamma_{p_x}, 0)
            \end{bmatrix}
           \end{gathered}$
       \\[5em] \toprule
\end{tabular}
    \caption{
        \captiontitle{Basis functions of total irreducible representations.}
        Allowed combinations of the spatial and spin irreps (\textit{total} irreps).
        }
    \label{tab:total_irreps}
\end{table}

\subsection{Irrep basis functions in the spin representation}
\label{app:spin_irreps}

To obtain the irrep of a spin pair, we begin with the symmetry group representation of a single spin ($S(\sym C {6v})$), which are the Euler rotations in Eqn.\,\eqref{eqn:euler_spin}.
Explicitly, the two generators are a $C_6$ rotation
\begin{equation}
    S(C_6) = \hat{U} \bigg( \frac{2 \pi}{6}, \hat{\bvec z} \bigg) = \frac{1}{2}
    \begin{pmatrix}
        \sqrt{3} - i & 0 \\
        0 & \sqrt{3} + i
    \end{pmatrix}\, ,
\end{equation}
when the lattice lies in the $xy$-plane, and a mirror plane $\sigma_1$, which is the composition of inversion and an in-plane rotation $C^{\prime}_2$. Since the spin is invariant under inversion, the mirror plane acts the same as the rotation $C^{\prime}_2$:
\begin{equation}
    S( \sigma_1 ) =  \hat{U} ( \pi, \hat{\bvec x} ) =
    \begin{pmatrix}
        0 & -i \\
        -i & 0
    \end{pmatrix}
    \, ,
\end{equation}
where we have specified that a bond has been aligned along the $x$-axis.
From these two representations, we can see that the five rotations around the principal $z$ axis do not flip spins, whereas the six mirror planes (\ie the in-plane rotations) do flip spins.

From the representations acting on a single spin $S(\sym C {6v})$ we obtain the
representation for two spins as the tensor product
$S(\sym C {6v}) \otimes S(\sym C {6v})$.
The basis functions for the spin pairs are shown in \cref{tab:irrep_basis_fns}, column three.

We can construct the basis functions for the total irrep from tensor products of
the spin-pairing and spatial pairing irreps.
We retain only the products that yield physical states, \ie those
which are antisymmetric under exchange of all quantum indices
(\eg \sym E 2 $\otimes$ \sym E 1 would yield spatially even
spin-triplet pairing and is discarded).
Except for the case $\sym E 1 \otimes \sym E 1$, either the spatial or the
spin-pair space is one-dimensional.
Since multiplication of the irreps' characters yields the total
state's characters, in these one-dimensional cases this is sufficient for determining
the total irrep.
The $\sym E 1 \otimes \sym E 1$ case requires a further decomposition, and
results in $\sym E 1 \otimes \sym E 1 = \sym A 1 \oplus \sym A 2 \oplus \sym E 2$.

We summarize all allowed combinations in \cref{tab:total_irreps}, as these correspond to the superconducting groundstates which are allowed by symmetry for the TLRHM. We stress that there are {\it three different} $E_2$ total irreps, which mix in different proportions the superconducting instabilities, resulting in different topological superconducting states.

\section{Harmonic Representation of Superconducting States}
\label{app:harmonics}

If the superconducting instability lies in the \sym E 2 irrep, the order parameter can take the form:
\begin{align}
    \Delta = \sum_{n=1}^3 \bigg(
        a_{d,n} & \begin{bmatrix} \Gamma^n_{d_{x^2-y^2}} \\ \Gamma^n_{d_{xy}}
        \end{bmatrix} +
        a_{f,n} \begin{bmatrix} (\Gamma^n_f, 0, 0) \\ (0, \Gamma^n_f, 0)
        \end{bmatrix}  \nonumber \\[10pt] & +
        a_{p,n} \begin{bmatrix} (\Gamma^n_{p_x}, -\Gamma^n_{p_y}, 0) \\
            (\Gamma^n_{p_y}, \Gamma^n_{p_x}, 0)
        \end{bmatrix}
        \bigg) \label{eqn:sc_inst_harm}
\end{align}
where we sum over the interaction shells $n=1,2,3$ corresponding to nearest-neighbor,
next nearest-neighbor and next-next nearest-neighbor Coooper pairing respectively.
There are 9 independent coefficients $a_{i,n}$.
In principle, nothing in the TUFRG flow or the basic group theoretical arguments prevents
these coefficients from being complex, however we find that while the
superconducting order parameter is only defined up to a an overall gauge,
the relative phases between the coefficients are real numbers.
The realness of these coefficients for a similar example for singlet-triplet mixed states under RSOC on the square lattice has been shown using Ginzburg-Landau theory
free energy minimization arguments\,\cite{holst2022}.

\begin{figure*}
    \includegraphics[width=\textwidth]{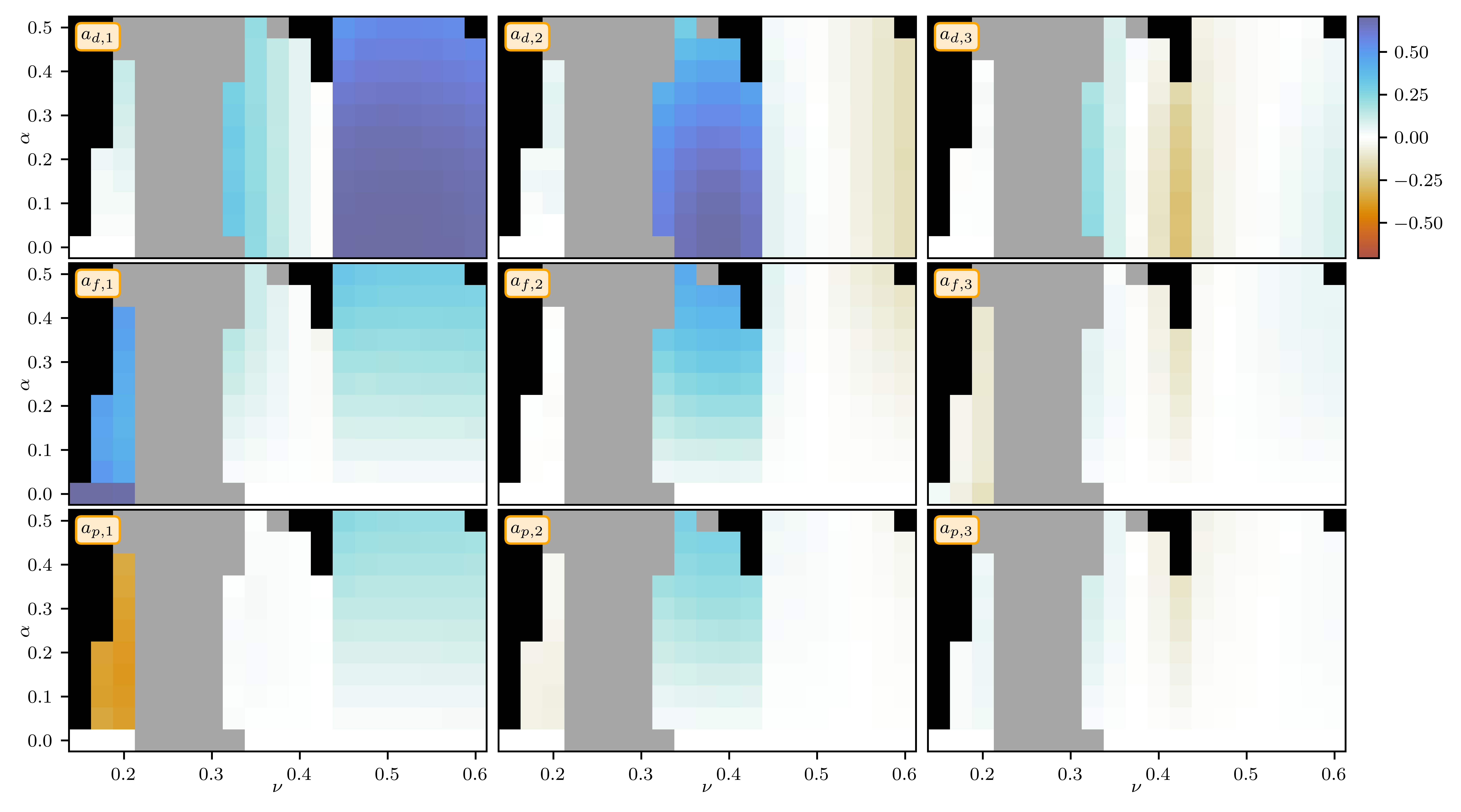}
    \caption{
        \captiontitle{Coefficients of the different basis functions for the \sym E 2 superconducting instability.} For each point in the phase diagram, the nine coefficients $a_{i,n}$ completely define the superconducting instability via Eqn.\,\eqref{eqn:sc_inst_harm} for the three types of \sym E 2 superconducting pairings $i \in \{ d, f, p \}$ on the first three interaction shells $n \in \{1,2,3 \}$. The forms of the harmonics can be read directly from Tab.\,\ref{tab:harmonics}. This information can be extracted directly from the TU formulation of our FRG method.
    }
    \label{fig:coeffs}
\end{figure*}

Harmonics can be explicitly figured out from the form of the irrep transformation
on each shell of form factors.
In form factor space, the functions are listed in the central column of Tab.\,\ref{tab:irrep_basis_fns},
calculated on the nearest neighbors, but applicable for form factor distances with 6 lattice sites.
We can express these functions in terms of trigonometric functions of the reciprocal momentum,
noting that in order for these functions to be real, the coefficients in \cref{tab:irrep_basis_fns}
are multiplied by $i$ to get real sine functions. These functions are also normalized. All these functions
are explicitly listed in \cref{tab:harmonics}.

Note also a subtlety of the analysis is that the second form factor shell
(next nearest-neighbors) is $\pi/3$ rotated to the other two shells. All the
same functions that are representable on the other shells can be represented on
this shell too, with the exception of the $\Gamma_f$, which is $f_{x(x^2-3y^2)}$
on the other shells. Instead, we can represent $f_{y(y^2-3x^2)}$, which is also
an $f$-wave pairing (and technically in $B_{2u}$, a different irrep of $D_{6h}$
than the nearest- and next-next-nearest-neighbor pairings which are $B_{1u}$, with a fixed orientation).
This difference means that we must swap the order of the $E_2$ irrep $f$-wave
pairing for this second shell when comparing to other states, \ie for $n=2$,
\begin{align}
    \Delta^{n=2} =
        a_{d,2} & \begin{bmatrix} \Gamma^2_{d_{x^2-y^2}} \\ \Gamma^2_{d_{xy}}
        \end{bmatrix} +
        a_{f,2} \begin{bmatrix} (0, \Gamma^2_f, 0) \\ (\Gamma^2_f, 0, 0)
        \end{bmatrix}  \nonumber \\[10pt] & +
        a_{p,2} \begin{bmatrix} (\Gamma^2_{p_x}, -\Gamma^2_{p_y}, 0) \\
            (\Gamma^2_{p_y}, \Gamma^2_{p_x}, 0)
        \end{bmatrix}
\end{align}

\begin{table}
    \centering
    \begin{tabular}{cc} \toprule
    & $n=1$ \\
    \hline\\[-5pt]
    $\Gamma^1_s$ & $\frac{\sqrt{6}}{3} \bigg( \cos(x)
    + \cos \bigg( \frac{ - x + \sqrt{3} y}{2} \bigg)
    + \cos \bigg( \frac{ - x - \sqrt{3} y}{2} \bigg)
    \bigg)$ \\
    $\Gamma^1_f$ & $\frac{\sqrt{6}}{3} \bigg( \sin(x)
    + \sin \bigg( \frac{ - x + \sqrt{3} y}{2} \bigg)
    + \sin \bigg( \frac{ - x - \sqrt{3} y}{2} \bigg)
    \bigg)$ \\
    $\Gamma^1_{p_x}$ & $\frac{\sqrt{3}}{3} \bigg( 2 \sin(x)
    - \sin \bigg( \frac{ - x + \sqrt{3} y}{2} \bigg)
    - \sin \bigg( \frac{ - x - \sqrt{3} y}{2} \bigg)
    \bigg)$ \\
    $\Gamma^1_{p_y}$ & $\bigg( \sin \bigg( \frac{ - x + \sqrt{3} y}{2} \bigg)
    - \sin \bigg( \frac{ - x - \sqrt{3} y}{2} \bigg)
    \bigg)$ \\
    $\Gamma^1_{d_{x^2-y^2}}$ & $\frac{\sqrt{3}}{3} \bigg( 2 \cos(x)
    - \cos \bigg( \frac{ x + \sqrt{3} y}{2} \bigg)
    - \cos \bigg( \frac{ x - \sqrt{3} y}{2} \bigg)
    \bigg)$ \\
    $\Gamma^1_{d_{xy}}$ & $\bigg( \cos \bigg( \frac{ - x + \sqrt{3} y}{2} \bigg)
    - \cos \bigg( \frac{ - x - \sqrt{3} y}{2} \bigg)
    \bigg)$ \\[10pt] \hline
    & $n=2$ \\
    \hline\\[-5pt]
    $\Gamma^2_s$ & $\frac{\sqrt{6}}{3} \bigg(
    \cos (y)
    + \cos \bigg( \frac{ 3 x - \sqrt{3} y}{2} \bigg)
    + \cos \bigg( \frac{ - 3 x - \sqrt{3} y}{2} \bigg)
    \bigg)$ \\
    $\Gamma^2_f$ & $\frac{\sqrt{6}}{3} \bigg(
    \sin (y)
    + \sin \bigg( \frac{ 3 x - \sqrt{3} y}{2} \bigg)
    + \sin \bigg( \frac{ - 3 x - \sqrt{3} y}{2} \bigg)
    \bigg)$ \\
    $\Gamma^2_{p_y}$ & $\frac{\sqrt{3}}{3} \bigg(
        2 \sin (y)
        - \sin \bigg( \frac{ 3 x - \sqrt{3} y}{2} \bigg)
        - \sin \bigg( \frac{ - 3 x - \sqrt{3} y}{2} \bigg)
    \bigg)$ \\
    $\Gamma^2_{p_x}$ & $\bigg(
        \sin \bigg( \frac{ 3 x - \sqrt{3} y}{2} \bigg)
        - \sin \bigg( \frac{ - 3 x - \sqrt{3} y}{2} \bigg)
    \bigg)$ \\
    $\Gamma^2_{d_{x^2-y^2}}$ & $\frac{\sqrt{3}}{3} \bigg(
        - 2 \cos (y)
        + \cos \bigg( \frac{ 3 x - \sqrt{3} y}{2} \bigg)
        + \cos \bigg( \frac{ - 3 x - \sqrt{3} y}{2} \bigg)
    \bigg)$ \\
    $\Gamma^2_{d_{xy}}$ & $\bigg(
        \cos \bigg( \frac{ 3 x - \sqrt{3} y}{2} \bigg)
        - \cos \bigg( \frac{ - 3 x - \sqrt{3} y}{2} \bigg)
    \bigg)$ \\[10pt] \hline
    & $n=3$ \\
    \hline\\[-5pt]
    $\Gamma^3_s$ & $\frac{\sqrt{6}}{3} \bigg( \cos(2x)
    + \cos ( - x + \sqrt{3} y ) + \cos ( - x - \sqrt{3} y ) \bigg)$ \\
    $\Gamma^3_f$ & $\frac{\sqrt{6}}{3} \bigg( \sin(2 x)
    + \sin (- x + \sqrt{3} y ) + \sin ( - x - \sqrt{3} y) \bigg)$ \\
    $\Gamma^3_{p_x}$ & $\frac{\sqrt{3}}{3} \bigg( 2 \sin(2 x)
    - \sin (- x + \sqrt{3} y ) - \sin ( - x - \sqrt{3} y) \bigg)$ \\
    $\Gamma^3_{p_y}$ & $\bigg(
    \sin (- x + \sqrt{3} y ) - \sin ( - x - \sqrt{3} y) \bigg)$ \\
    $\Gamma^3_{d_{x^2-y^2}}$ & $\frac{\sqrt{3}}{3} \bigg( 2 \cos(2 x)
    - \cos (- x + \sqrt{3} y ) - \cos ( - x - \sqrt{3} y) \bigg)$ \\
    $\Gamma^3_{d_{xy}}$ & $\bigg(
        \sin (- x + \sqrt{3} y ) - \sin ( - x - \sqrt{3} y)
    \bigg)$ \\[10pt] \hline
    \end{tabular}
    \caption{
        \captiontitle{Explicit functional forms of the harmonics in reciprocal space.} Together with the coefficients in Fig.\,\ref{fig:coeffs}, these functions can be used to completely define the superconducting instabilties via Eqn.\,\ref{eqn:sc_inst_harm}.
        }
    \label{tab:harmonics}
\end{table}

\section{Bogoliubov--de Gennes analysis}

\subsection{BdG Formalism}

If the TUFRG flow results in a divergence in the $P$-channel at $\bvec q = 0$, the low-energy effective interacting Hamiltonian is
\begin{align}
    H =& \sum_{\bvec{k}, ss'} h_{ss'} (\bvec{k}) c^{\dagger}_{\bvec{k} s} c^{\phantom{\dagger}}_{\bvec{k} s'} \nonumber \\
    &+ \frac{1}{2} \sum_{\bvec{k} \bvec{k}', ss'rr'} \!\!\!\! V_{ss'rr'} (\bvec{k}, \bvec{k}') c^{\dagger}_{\bvec{k} r} c^{\dagger}_{-\bvec{k} r'} c^{\phantom{\dagger}}_{-\bvec{k}' s'} c^{\phantom{\dagger}}_{\bvec{k}' s} \, ,
\end{align}
where $\hat h$ is the (kinetic) Bloch Hamiltonian and $V_{ss'rr'} ( \bvec{k}, \bvec{k}') = V_{ss'rr'}^P (\bvec{q} = 0, \bvec{k}, \bvec{k}') $ is the effective superconducting vertex. \\
By introducing a mean-field approximation
\begin{equation}
    c_{-\bvec k s} c_{\bvec k s'} = \langle c_{-\bvec k s} c_{\bvec k s'} \rangle + ( c_{-\bvec k s} c_{\bvec k s'} -
    \langle c_{-\bvec k s} c_{\bvec k s'} \rangle) \, ,
\end{equation}
and assuming fluctuations are small, so the product of two fluctuation terms is negligible,
the quartic interation can be reduced to a quadratic form
\begin{align}
    H = \sum_{\bvec{k}, ss'} &h_{ss'} (\bvec{k}) c^{\dagger}_{\bvec{k} s} c^{\phantom{\dagger}}_{\bvec{k} s'}
    + \mathcal{K} \nonumber \\
    &+ \frac{1}{2} \sum_{\bvec{k}, s s'}
    \Delta^{\phantom{*}}_{s s'} (\bvec{k})  c^{\dagger}_{\bvec{k} s} c^{\dagger}_{-\bvec{k} s'} + \text{h.c.} \, ,
\end{align}
where we define
\begin{align}
    \Delta_{s s'} (\bvec k) &= \frac{1}{2} \sum_{\bvec k', rr'}
    V_{rr'ss'} (\bvec k, \bvec k') \; \langle c_{-\bvec k' r'} c_{\bvec k' r} \rangle\ , \\
    \mathcal{K} &= -\frac{1}{2} \sum_{\bvec k \bvec k', ss'rr'} \!\!\!\! V_{ss'rr'} (\bvec k, \bvec k')
    \langle c^{\dagger}_{\bvec k r} c^{\dagger}_{-\bvec k r'} \rangle \langle c^{\phantom{\dagger}}_{-\bvec k' s'} c^{\phantom{\dagger}}_{\bvec k' s} \rangle .
\end{align}
We now perform the canonical BdG transformation, and can write this in the form:
\begin{equation}
    H = \frac{1}{2} \sum_{\bvec{k}} C^{\dagger}_{\bvec{k}} \hat{H}_{\text{BdG}} (\bvec{k}) C^{\phantom{\dagger}}_{\bvec{k}} + \frac{1}{2} \sum_{\bvec k} \text{Tr} (\hat{h} (\bvec k)) + \mathcal{K}
    \end{equation}
    with spinor
    \begin{equation}
    C_{\bvec{k}} = \begin{pmatrix} c_{\bvec{k} \downarrow} & c_{\bvec{k} \uparrow} &
        c^{\dagger}_{-\bvec{k} \downarrow} & c^{\dagger}_{-\bvec{k} \downarrow} \end{pmatrix}^T \, ,
        \end{equation}
        and the BdG matrix
        \begin{equation}
    \hat{H}_{\text{BdG}} = \begin{pmatrix} \hat{h} (\bvec{k}) & \hat{\Delta} (\bvec{k}) \\
        \hat{\Delta}^\dagger (\bvec{k}) & - \hat{h}^T (-\bvec{k}) \end{pmatrix}\ .
\end{equation}
%
%
%
%
We can then diagonalize the BdG matrix, forming Bogoliubov quasiparticles $\alpha_{\bvec{k} r}$,
with energy $E_{\bvec{k} r}$, which are free particles that follow Fermi statistics.
\begin{equation}
    H = \frac{1}{2} \sum_{\bvec{k}} A^{\dagger}_{\bvec{k}} \hat{E} (\bvec{k}) A^{\phantom{\dagger}}_{\bvec{k}} + \frac{1}{2} \sum_{\bvec k} \text{Tr} (\hat{h} (\bvec k)) + \mathcal{K}
    \end{equation}
    with spinor
    \begin{equation}
    A_{\bvec{k}} = \begin{pmatrix} \alpha_{\bvec{k} \downarrow} & \alpha_{\bvec{k} \uparrow} &
        \alpha^{\dagger}_{-\bvec{k} \downarrow} & \alpha^{\dagger}_{-\bvec{k} \downarrow} \end{pmatrix}^T
        \end{equation}
        and
        \begin{equation}
    \hat{E} (\bvec{k}) = \text{diag} \begin{pmatrix} E_0 (\bvec{k}) & E_1 (\bvec{k}) &
        - E_0 ( - \bvec{k}) & - E_1 (- \bvec{k}) \end{pmatrix}\ .
\end{equation}
The two bases are related by a unitary (Bogoliubov) transformation:
\begin{align}
    \hat{H}_{\text{BdG}} &= \hat{U} \hat{E} \hat{U}^\dagger \\
    \hat{U}& = \begin{pmatrix}
        \hat{u}_{\bvec{k}} & \hat{v}^*_{-\bvec{k}} \\
        \hat{v}_{\bvec{k}} & \hat{u}^*_{-\bvec{k}}
    \end{pmatrix} \\
    C_{\bvec{k}} &= \hat{U} A_{\bvec{k}}
    \end{align}
    which corresponds to the more familiar expressions
    \begin{align} &\begin{cases}
        c_{\bvec{k} s} = u_{\bvec{k} ss'} \alpha_{\bvec{k} s'} + v^{*}_{-\bvec{k} ss'} \alpha^{\dagger}_{-\bvec{k} s'} \\[5pt]
        c^{\dagger}_{\bvec{k} s} = u^*_{\bvec{k} ss'} \alpha^\dagger_{\bvec{k} s'} + v_{-\bvec{k} ss'} \alpha_{-\bvec{k} s'}\ .
    \end{cases} \label{eqn:bogoliubov_expansion}
\end{align}
The top line above tells us that the columns of $\hat{U}$ are the eigenvectors of
$\hat{H}_{\text{BdG}}$.

\subsection{Doubly degenerate SC order parameter and free energy minimization}

In general, we can form an arbitrary complex superposition of the two degenerate superconducting instabilities,
\begin{equation}
    \Delta = \cos (\theta) \Delta_1 + e^{i \varphi} \sin (\theta) \Delta_2\ .
\end{equation}
The stable superposition of SC order parameter can be determined by minimizing the free energy of the BdG Hamiltonian at critical scale $\Lambda$\,\cite{klebl2023, platt2013}, which is given as
\begin{equation}
    F = \langle H \rangle - \Lambda S\ . \label{eqn:free-en}
    \end{equation}
Here the internal energy is
    \begin{align}
    \langle H \rangle &= \sum_{\bvec{k}, n} E_n (\bvec{k}) \left(f \bigg( \frac{E_n (\bvec{k})}{\Lambda} \bigg) - \frac{1}{2} \right)
    + \frac{1}{2} \sum_{\bvec{k}, s} h_{ss} (\bvec{k}) \nonumber \\
    &+ \frac{1}{2} \sum_{\bvec{k}, s s' n}  \Delta_{s s'}(\bvec{k}) u^*_{\bvec{k} s n} v_{\bvec{k} s' n}
    \, \text{tanh} \bigg( \frac{E_n (\bvec{k})}{2 \Lambda} \bigg)\
    \end{align}
    and the entropy
    \begin{align}
    S &= \sum_{\bvec{k}, n} \bigg[ f \bigg( \frac{E_n (\bvec{k})}{\Lambda} \bigg) \log \bigg( f \bigg( \frac{E_n (\bvec{k})}{\Lambda} \bigg) \bigg) \nonumber \\
    &\hspace{1cm} + f \bigg( -\frac{E_n (\bvec{k})}{\Lambda} \bigg) \log \bigg( f \bigg( -\frac{E_n (\bvec{k})}{\Lambda} \bigg) \bigg) \bigg]
\end{align}
where $f(x) = (e^x + 1)^{-1}$ is the Fermi function. We sum over spin indices $s$ and BdG band indices $n$, for the two positive BdG bands.

\begin{figure}[t!]
    \centering
    \includegraphics[width=0.44\textwidth]{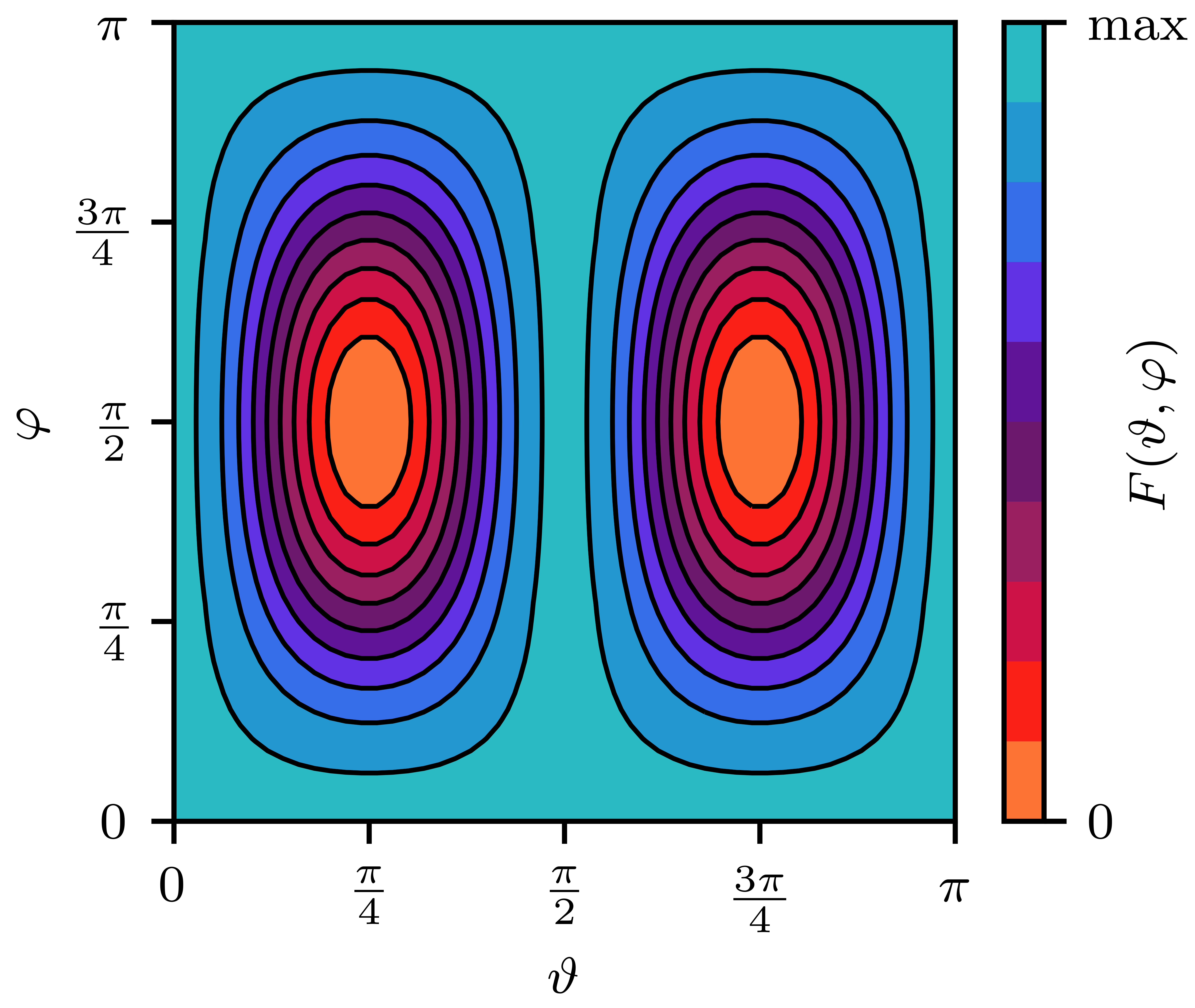}
    \caption{
        \captiontitle{Free energy variation with complex superpositions of superconudcting pairings.} Eq.\,\eqref{eqn:free-en}, calculated for the superconducting instability at $\nu = 0.4$, $\alpha = 0.3$. The stable superpositon of the superconductor is the minima of the free energy (in arbitrary units), which are the chiral superpositions $\Delta = (\Delta_1 \pm i \Delta_2) / \sqrt{2}$.
        }
    \label{fig:free_en}
\end{figure}

Our free energy calculations reveal that the complex equal-weight, i.e., chiral, superposition is always energetically favored. What is expected in the absence of RSOC $\alpha$ persists to finite spin-orbit coupling.  The low-filling phase at $\alpha=0$ is an exception as the spatial irrep is one-dimensional ($B_1$). A representative example of the free energy showing chiral superconductivity for $\alpha=0.3$ at filling $\nu=0.3$ is presented in Fig.\,\ref{fig:free_en}.

\subsection{Topological Analysis}

\subsubsection{Chern number}

Superconducting phases can be characterized by a topological invariant, \ie the Chern number in the BdG gap, which is the sum of the Chern numbers of the lower two bands:
\begin{equation}
    \mathcal C = \frac{1}{2 \pi i} \sum_{n=1}^2 \iint_{\text{BZ}} \dd^2{\bvec k} \, \bigg( \frac{\partial A^{(n)}_y}{\partial k_x} - \frac{\partial A^{(n)}_x}{\partial k_y} \bigg)
\end{equation}
where ${\bvec A}^{(n)} (\bvec k) = \langle n (\bvec k) | \nabla_{\bvec k} | n (\bvec k) \rangle $ is the Berry connection on the $n$th BdG band, and $| n(\bvec k) \rangle$ the $n$th eigenstate of the BdG Hamiltonian. This integral can be efficiently computed using numerical methods\,\cite{fukui2005}, as long as the superconducting phase is gapped.
We note that other works consider the Chern number due to only one of the doubly degenerate bands when classifying gapped superconducting phases, however when RSOC is introduced, the spin degeneracy of the BdG bands is lifted, so to remain consistent over the entire phase diagram, we calculate the Chern number as the sum over the two lower BdG bands.

\subsubsection{Ribbon spectra}

Due to the bulk-boundary correspondence, the Chern number of the superconducting phase can be determined by the number of edge states which traverse the bulk in a ribbon strip geometry\cite{laughlin1981}. The ribbon geometry is a strip of $N$ lattice sites with open boundary conditions in the $\bvec{a}_1$, and we Fourier transform along the periodic $\bvec{a}_2$ direction.

To map the superconducting instability expressed in terms of form factors to the ribbon geometry, we map the form factors onto the grid of lattice vectors $\bvec f = n \, \bvec{a}_1 + m \, \bvec{a}_2$. Then, the SC order parameter for the pairing between one lattice site and another $-F \leq n \leq F, \, F=\text{max}(|\bvec{f}|)$ along the strip is given by
\begin{equation}
    \hat{\Delta}_n (k) = \sum_{m} e^{- i m k} \hat{\Delta}_{(n, m)}  \, ,
\end{equation}
where $\hat{\Delta}_{(n,m)} = \hat{\Delta}_f$ is the SC order parameter in form factor space, with the hat signifying that the order parameter is a matrix in spin space. The full SC order parameter is given by
\begin{equation}
    \hat{\Delta}_{i,j} = \hat{\Delta}_{j-i} \, ,
\end{equation}
with $1 \leq i,j \leq N$. The full SC order parameter is then a matrix with dimensions $2N \times 2N$. This is then used in the BdG Hamiltonian, which is diagonalized to get the BdG ribbon spectrum. Edge states, such as those in Fig.\,2, are identified by BdG eigenvectors where greater than 85\% of the spectral weight lies on one side of the strip, with `left' states (\ie states where $>85$\% of the spectral weight lies on sites $1 \leq i \leq N/2$) colored blue, `right' states ($N/2 < i \leq N$) colored red, and bulk states in black.
Since the SC amplitude $|\Delta|$ is of the order of $~ 10^{-3} t$, compared to a bandwidth of $~ 10 t$, large ribbon strip sizes are needed to resolve the superconducting gap and any edge states.

\begin{figure}[t!]
    \centering
    \includegraphics[width=0.22\textwidth]{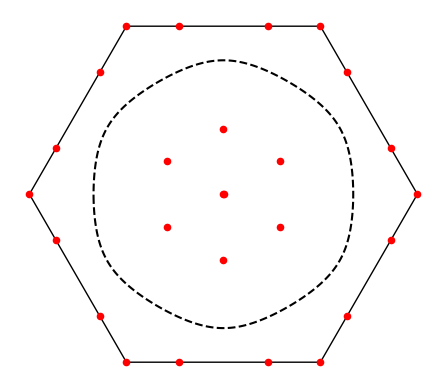}
    \includegraphics[width=0.22\textwidth]{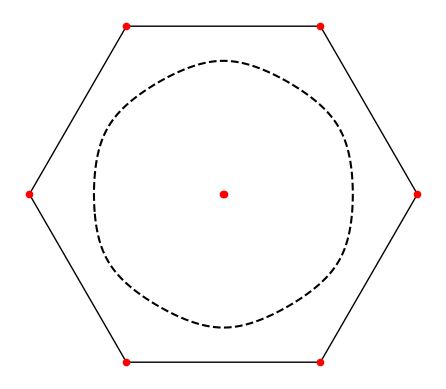}
    \caption{
        \captiontitle{Nodal points in the superconducting pairing.} Black solid line is the Brillouin Zone boundary, dashed line is the Fermi surface, and the red dots are the nodes in the superconducting pairing. (Left) Filling $n=0.42$, Chern number $C=-8$  (Right) Filling $n=0.46$, Chern number $C=4$
        }
    \label{fig:nodal}
\end{figure}

\subsection{Topological phase transition with no Rashba SOC}

In the case of no RSOC ($\alpha = 0$), and unitary SC pairing, \textit{i.e.}, $\hat{\Delta} = \hat{\Delta}^{\dagger}$,
we know the analytic form of the BdG energies $E_0 = E_1 = E$:
\begin{equation} \label{eqn:bdg_ana}
    E = \sqrt{ (\varepsilon_0 (\bvec{k}))^2 + |\Delta (\bvec k)|^2} \, ,
\end{equation}
where $\varepsilon_0 (\bvec k)$ is the (normal state) dispersion relation, as defined in the main text, and $|\Delta (\bvec k)|$ is the magnitude of the order parameter,
\begin{equation}
    |\Delta (\bvec k)| = \sqrt{\frac{1}{2} \text{Tr} [\hat{\Delta}^\dagger (\bvec k) \hat{\Delta} (\bvec k) ]} \, .
\end{equation}
From Eqn.\,\eqref{eqn:bdg_ana}, we can see that gap closings in the BdG spectrum -- and therefore topological phase transitions -- can only occur when the Fermi surface intersects the zeros of the superconducting pairing. When there is a chiral superposition of the doubly degenerate $d$-wave spin singlet SC pairing, \ie, when
\begin{equation}
    \hat{\Delta} (\bvec k) = (\Gamma_{d_{x^2 - y^2}} + i \, \Gamma_{d_{xy}} ) ( i \hat{\sigma}^y ) \, ,
\end{equation}
the zeros of this SC pairing are points - or vortices - in the Brillouin Zone, as opposed to lines. The location of the vortices is determined by distribution of the SC order parameter weight across the different interaction shell lengths, Fig.\,\ref{fig:nodal} shows the location of these vortices for two points on either side of the phase transition at $\nu_c = 0.445$.

We can ascribe a topological charge to each of these vortices, which is the winding number of the complex phase of $\Delta_{\uparrow \downarrow} (\bvec k)$ in the neighborhood of the vortices. The Chern number of the band is then the winding number of $\Delta_{\uparrow \downarrow} (\bvec k)$  on the Fermi surface, which is the sum of vortices contained within the FS. The vortex at the $\Gamma$ point has a winding number of $+2$ (which is expected of a $d$-wave pairing), and each of the vortices within the FS not at the $\Gamma$ point has a charge of $-1$.
Therefore, the Chern number for one band is $\mathcal C = + 2$ when the only vortex within the FS is at the $\Gamma$ point, as in Fig.\,\ref{fig:nodal} (b) and $\nu > \nu_c$, and $\mathcal C = -4$ when there are 6 other symmetry-related vortices bound by the FS, as in Fig.\,\ref{fig:nodal} (a) and $\nu < \nu_c$. The Chern number of the gapped SC phases are then $\mathcal C = 4, -8$ respectively, as we sum over the two degenerate bands.





\bibliography{refs_TLRHM}